\documentclass[preprint]{elsarticle}

\usepackage{amsmath,amssymb,amsfonts}
\usepackage{algorithmic}
\usepackage{graphicx}
\usepackage{booktabs}
\usepackage{tabularx}
\usepackage{multirow}
\usepackage{siunitx}
\usepackage{textcomp}
\usepackage{xcolor}
\usepackage{hyperref}
\usepackage{xurl}
\usepackage{tcolorbox}
\usepackage{pifont}
\usepackage{balance}
\usepackage{listings}

\definecolor{delim}{RGB}{20,105,176}
\definecolor{numb}{RGB}{106, 109, 32}
\definecolor{string}{rgb}{0.64,0.08,0.08}
\lstdefinelanguage{json}{
    frame=single,
    rulecolor=\color{black},
    showspaces=false,
    showtabs=false,
    breaklines=true,
    postbreak=\raisebox{0ex}[0ex][0ex]{\ensuremath{\color{gray}\hookrightarrow\space}},
    breakatwhitespace=true,
    basicstyle=\ttfamily\small,
    upquote=true,
    morestring=[b]",
    stringstyle=\color{string},
    literate=
     *{0}{{{\color{numb}0}}}{1}
      {1}{{{\color{numb}1}}}{1}
      {2}{{{\color{numb}2}}}{1}
      {3}{{{\color{numb}3}}}{1}
      {4}{{{\color{numb}4}}}{1}
      {5}{{{\color{numb}5}}}{1}
      {6}{{{\color{numb}6}}}{1}
      {7}{{{\color{numb}7}}}{1}
      {8}{{{\color{numb}8}}}{1}
      {9}{{{\color{numb}9}}}{1}
      {\{}{{{\color{delim}{\{}}}}{1}
      {\}}{{{\color{delim}{\}}}}}{1}
      {[}{{{\color{delim}{[}}}}{1}
      {]}{{{\color{delim}{]}}}}{1},
}

\def\BibTeX{{\rm B\kern-.05em{\sc i\kern-.025em b}\kern-.08em
    T\kern-.1667em\lower.7ex\hbox{E}\kern-.125emX}}

\begin{document}

\begin{frontmatter}



\title{An Allocation Model for Attributing Emissions in Multi-tenant Cloud Data Centers}


\author[inst1]{Richard Westerhof}

\affiliation[inst1]{organization={Department of Computer Science, University of Groningen}, \\
            addressline={Nijenborgh~9}, 
            city={Groningen},
            postcode={9474 AG}, 
            state={Groningen},
            country={the Netherlands}}

\author[inst2]{Richard Atherton}
\author[inst1]{Vasilios Andrikopoulos}

\affiliation[inst2]{organization={BT Global Services},
            addressline={Herikerbergweg 2}, 
            city={Amsterdam},
            postcode={1101 CM}, 
            state={Noord Holland},
            country={the Netherlands}}

\begin{abstract}
Cloud computing has become the de facto paradigm for delivering software to system users, with organizations and enterprises of all sizes making use of cloud services in some way. 
On the surface, adopting the cloud appears to be a very efficient approach for offloading concerns such as infrastructure management, logistics, and most importantly for this work, energy consumption and consequent carbon emissions to the cloud service provider.
However, this is in many ways not an appropriately accountable solution to managing the contribution of the ICT sector to global emissions.
To this effect, in this paper we report on an exploratory case study done in collaboration with a Software as a Service provider operating globally in the telecommunications sector. 
The study reckons with the service provider using multi-tenant, that is, shared, off-premises data centers for hosting their private cloud infrastructure towards developing a fair model of allocating operational emissions among the service tenants --- customer companies with many distinct users.

The developed emissions model has to account for allocating in an appropriate manner the generated emissions between the tenants of the software provider services, and among the different tenants of the same data center.
A carbon footprint report generator is developed building on the proposed model which is, in turn, used to present sustainability reports to involved stakeholders for evaluation purposes.
Our results show that the model is perceived as transparent, informative, and fair, with requested improvements focusing mainly on the generated reports and the information contained therein.
\end{abstract}



\begin{keyword}
cloud computing \sep carbon footprint \sep data centers \sep private cloud \sep software as a service \sep multi-tenancy \sep emissions modelling
\end{keyword}

\end{frontmatter}


\section{Introduction}
\label{sec:introduction}

Since its emergence, cloud computing has been adopted by enterprises, governmental organizations and the academia and has seen a rapid growth~\cite{buyya2018manifesto}.
This growth is the result of offering considerable advantages in rapidly provisioning on-demand access to all kinds of computational resources in the most cost efficient and reliable manner~\cite{armbrust2010view}.
The study of cloud computing therefore naturally forms an intersection as a research topic with that of sustainability, with many works appearing over the years as summarized in a series of surveys.
As early as 2011, for example, Beloglazov et al.~\cite{beloglazov2011} reviewed several works on this topic, focusing on application design, energy management, and virtualization. 
Mastelic et al.~\cite{mastelic2014} and Moghaddam et al.~\cite{moghaddam2015} performed similar surveys a few years later, with the survey of Gill and Buyya~\cite{gill2018} in 2018 being the latest and most exhaustive one on the same topic.
Even books such as the ones by Bitterlin~\cite{greenit2012} and Smith~\cite{smith2014} have already been written on the subject.
All that is not even accounting for the efforts in green computing as summarized for example in~\cite{kumar-2012}, which forms a parallel discussion on the same topic.

Nevertheless, in recent years the environmental impact of IT in general, and cloud computing in particular has started coming into sharp focus.
As a reaction to this, many organizations have set becoming carbon neutral, or even negative in the case of Microsoft\footnote{\url{https://blogs.microsoft.com/blog/2020/01/16/microsoft-will-be-carbon-negative-by-2030/} (Accessed May 8, 2023).}, as a goal for the following years.
Since most gross (physical) emissions cannot be reduced to zero, at least not with the currently available technologies, the goal therefore becomes to reduce net (compensated through other means) emissions to zero instead.
The latter is achieved through e.g.\ the direct use of renewable energy sources and providers, or through the purchasing of Renewable Energy Certificates (RECs)~\cite{rec}.
In either case, the first step towards offsetting all produced emissions is determining how many emissions were actually generated.

Serving this need, all three of the big cloud service providers (Amazon Web Services, Google Cloud Platform, and Microsoft Azure), collectively known as the ``hyperscalers'' because of their capacity to scale virtually with any demand, are providing their customers with carbon footprint monitoring dashboards.
Access to these dashboards and their emissions calculations is of course limited to customers of these companies, making them unsuitable for e.g.\ private cloud deployments, where the cloud infrastructure is provisioned for use by a single organization and is therefore not hosted in the hyperscalers' data centers.
Furthermore, even if the emissions calculation is the most important matter at hand, to inverse Mytton's argument in~\cite{mytton2020hiding}, the underlying data to independently verify the calculations provided by these tools is simply not available.

Developing any kind of model for calculating emissions in the cloud, however, faces two major challenges.
The first is the lack of insights and data into the data centers powering the cloud, even for organizations that own, manage and operate their own data centers. 
Instrumenting for example all devices in a data center and measuring all metrics required by the approaches proposed in the literature is in many cases unrealistic or not cost efficient since it requires manual labor to retrieve this data.
More importantly, correctly measuring and attributing emissions to each involved organization becomes much more challenging due to the inherent multi-tenant nature of the resource pooling employed by the providers, which is an essential characteristic of cloud computing~\cite{nistcloud2011}.
Multi-tenancy is the concept that cloud infrastructure, in terms of both physical and virtual resources, is shared among the users of the same cloud service.
Calculating emissions in this case becomes an exercise of assigning responsibility for the cumulative emissions and identifying an appropriate manner to distribute them among all users of the same cloud service across potentially multiple data centers.

This difficulty is only exacerbated by the fact that the guidance for applying the Greenhouse Gas (GHG) Protocol~\cite{ghgprotocol}, the de facto standard for reporting emissions across industries today, in the ICT industry~\cite{ghgguidance} requires data that are not easily available, especially for public cloud providers~\cite{mytton2020assessing}. 
Even if these data are available, the guidance itself does not include any support for measuring and reporting the emissions of providers serving multi-tenant software to their customers while using shared, i.e.\ multi-tenant data centers for doing so.
The state of the art literature as summarized later in Section~\ref{sec:related} is not much help either, focusing on data center operators as a whole and not distinguishing the emissions at tenant level.
As such, we need to look further into developing an appropriate approach for this purpose.

The main research question that this study therefore aims to address can be summarized as follows:
\begin{quote}
    \itshape How can the Total Carbon Footprint (TCF) of operational emissions be accurately and fairly attributed to each tenant in a multi-tenant data center?
\end{quote}
In order to answer this question we decompose it first into the following research questions:
\begin{enumerate}
    \item[\bfseries RQ1] How to apply the GHG Protocol to report the TCF of tenants in multi-tenant data centers?
    \item[\bfseries RQ2] How can the TCF of a data center be accurately obtained?
    \item[\bfseries RQ3] How can the TCF be fairly distributed over all tenants of a multi-tenant data center?
\end{enumerate}
To keep the scope of this work manageable we limit it to operational emissions, leaving the inclusion of embodied emissions~\cite{rodrigues2010carbon}, that is emissions due to the manufacturing, transportation, and disposal of data center equipment as future work.
We opt to address these questions in this scope by means of an exploratory case study~\cite{runeson2009} conducted in collaboration with an industrial partner that acts as a Software as a Service (SaaS) provider in the telecommunications sector serving multiple multinational companies.
This allows us to ground our investigation to the reality of actual practice, while offering us access to empirical data directly from the source (in this case, data centers) that otherwise we would not have.

In the process of answering the research questions identified above we develop a TCF calculation model which combines measured and estimated data into per tenant attributable emissions in a given time period, and a report generator based on this model that we use to evaluate the efficacy of our proposal.
These two items are the main contributions of this work and are discussed in Sections~\ref{sec:model} and~\ref{sec:evaluation}, respectively, after the case is explained in more detail in Section~\ref{sec:case}.
Some of the necessary background is covered in Section~\ref{sec:background}, related works are summarized in Section~\ref{sec:related}, and finally Section~\ref{sec:conclusions} concludes this paper and discusses some future work.


%

\section{Background}
\label{sec:background}

As discussed in the previous section, the GHG (Greenhouse Gas) Protocol is the de facto standard for reporting emissions across industrial sectors.
It distinguishes emissions in three scopes based on how much of a direct consequence of an organization's activities are the emissions:
\begin{itemize}
    \item Scope 1 includes all emissions which are a direct result of an organization's operations. This includes, for example, the operating of furnaces, fuel-burning machines, or generators. Beyond using emergency electricity generators, data centers have usually very little Scope 1 emissions.
    \item Scope 2 includes all emissions caused by the generation of electricity used by the devices of an organization. In the case of a data center, this entails servers, network routers and switches, cooling and lighting systems, etc. This scope is likely to contain most emissions caused by data centers and, as such, attributable to the data center owner rather than the tenant using any service hosted in the data center. In the guidance paper provider by Sotos~\cite{sotos2015}, organizations are encouraged to report energy consumption of  \emph{any} form under this scope.
    \item Scope 3 is reserved for all other emissions not reported in the other scopes, and is labeled optional by the GHG Protocol, in the sense that they are not always easy to be measured. However, these emissions can make a large difference in the actual footprint of an organization and should also be acknowledged, even though this varies by the sector of operations~\cite{matthews2008}. Similar to the guidance work for Scope~2, Bhatia and Ranganatham~\cite{bahtia2011} have written a guidance for Scope~3 emissions, separating them into two categories: upstream and downstream. Upstream include purchased goods and services, transportation and distribution, waste generated by operations, and business travel, while downstream include processing of sold products, end-of-life treatment of products, franchises, and investments.
\end{itemize}
Matthews et al.~\cite{matthews2008} also proposed Scope~4 as a way of decluttering Scope~3 emissions reporting, but this proposal has not been widely accepted at the time of writing.
An important point to take into consideration here is that the scope is different depending on whether we are using the data center operator or tenant perspective.
For example, energy consumed by server devices are in Scope~2 for the operator, since they purchased the energy, but Scope~3 for the data center tenant, since it is an emission indirectly caused through the consumption of the services offered by the data center.

Another interesting point is raised by the work of Lenzen and Murray~\cite{lenzen2010} discussing upstream and downstream indirect emissions.
They basically argue that any entity that buys a product or service is at least partially responsible for the entire upstream emissions. 
Similarly, any entity that sells a product or service is (partially) responsible for the entire downstream emissions. 
The further up or down the stream emissions reside, the less responsible an entity is. 
Therefore, calculating the total emissions is an infinite sum with a finite result.
This observation builds on the prior work of Lenzen et al.~\cite{lenzen2007} that showed that Life Cycle Assessment (LCA) calculations only work when the emission responsibilities are pushed all the way to either end of the production life cycle.
However, this is fundamentally unfair, and responsibility should be shared between the producer and consumer of a product or service.
For this purpose, Lenzen et al.\ propose to define the shared responsibility as a set percentage between these two parties: if the producer is responsible for $x\%$ of the emissions, the consumer is responsible for the remaining $1-x\%$.
This is a proposal that we will be using later in this work.

\section{Case Design}
\label{sec:case}

In the following we elaborate on the details of the case study we performed as the means of answering the research questions defined in the introductory section.

\subsection{Case Description}

BT Global Services is a division of the BT Group (formerly known as British Telecom) focusing on providing security, cloud, and networking services to multinational companies worldwide\footnote{In the following we will be using simply BT as a shorthand for BT Global Services. This is not to reflect the whole of BT as a company.}.
One of the services offered to their customers is Cloud Contact Cisco (CCC), a product developed by Cisco providing virtualized call center capabilities in the SaaS model. 
BT's customers use this software for their customer service departments. 
The end users of CCC are the employees of the customer companies that subscribe to BT's service.
These end users are internally referred to as \emph{agents}.
Each service tenant (customer company) has multiple agents using the software at any point in time, with the amount of concurrent agents varying per tenant as different time zones enter and leave working hours.

\begin{figure}
    \centering
    \includegraphics[width=\columnwidth]{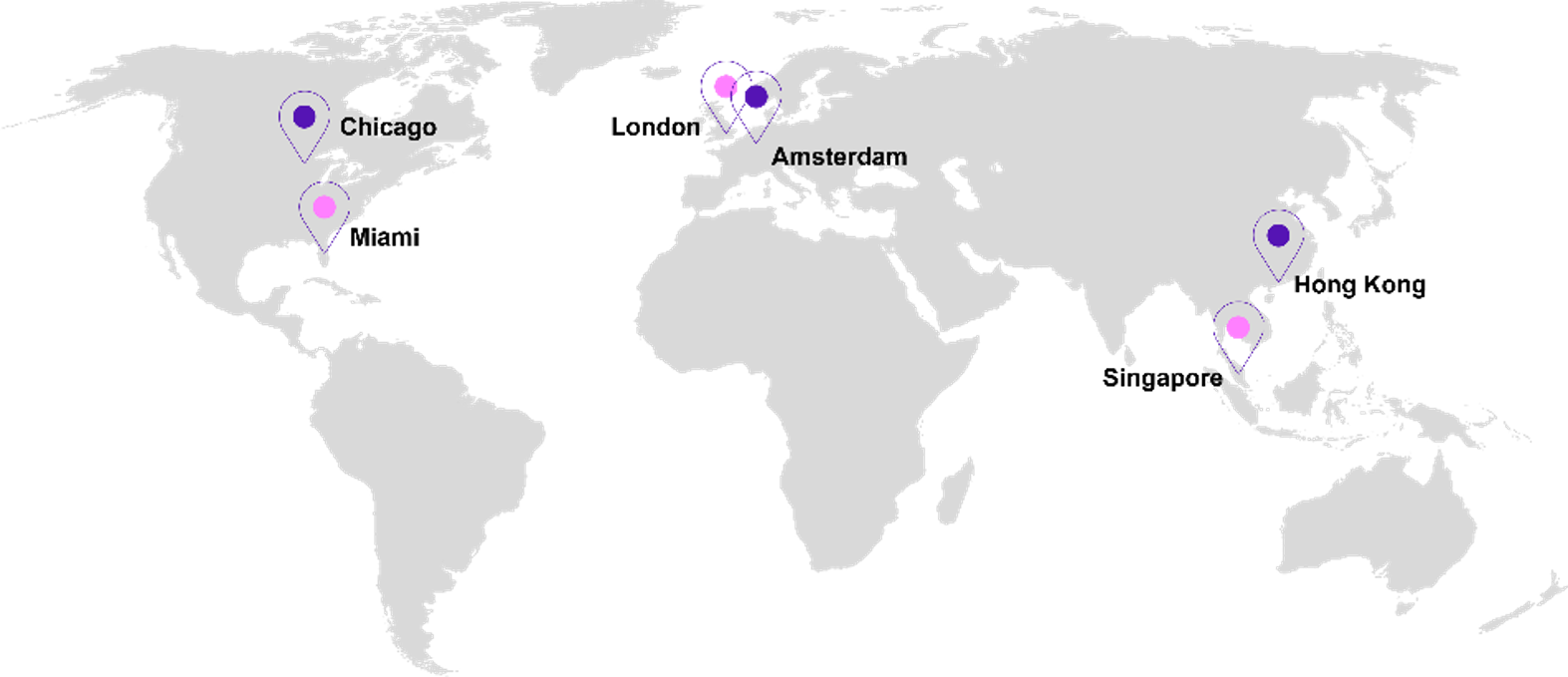}
    \caption{The distribution of the data centers relevant for this case study across the globe.}
    \label{fig:DCs}
\end{figure}

The software itself is deployed in several data centers across the globe as shown in Fig.~\ref{fig:DCs}, with the computing and storage infrastructure (i.e.~server racks and storage solutions) being exclusively assigned to service tenants, but with the network infrastructure being shared among them.
Each data center hosts multiple software solutions for BT beyond CCC operating as an off-premises private cloud, as well as software solutions of other companies, and is managed by BT.
As such, energy consumption for cooling, lighting, etc.\ is effectively shared among the different products that BT is hosting in each such center.
Furthermore, each data center comes with its own set of capabilities and limitations with respect to what can be monitored and shared.
A different version of the same software is also available through a deployment in one of the hyperscalers' public cloud services instead of the private cloud instance.
However, at the time of writing this, the majority of BT's customers are exclusively using the private cloud solution.

One more important aspect of the case is the existing, ongoing efforts of BT and the data center operating companies it employs to address emission issues.
With respect to the latter, electricity providers with a significant or complete proportion of renewable sources in their mix are being selected for the operation of the data centers.
This is combined with the data centers e.g.\ installing solar panels and wind generators as additional renewable sources of electricity.
BT, on the other hand, complements these efforts by purchasing RECs as the means to ensure that an as large as possible portion of the operational emissions is actually compensated.

With all that in mind, in the following we define the problem to be addressed by this study.

\subsection{Problem Definition}

In recent years, BT has taken notice of their customers becoming increasingly aware of their environmental impact, and requiring better insights into their footprint with respect to using BT's services such as CCC.
With the hyperscalers already offering their customers the capability to monitor their carbon footprint in the cloud, as discussed in Section~\ref{sec:introduction}, and with the majority of BT's customers still not having opted for the public cloud version of CCC, it falls to BT to provide them with equivalent capabilities.
For this purpose, they have developed in cooperation with a third party a carbon footprint dashboard which feeds on power meters installed at the used data centers.
The issue with this solution is that, as it is explained in the previous section, it is not possible to install such meter at each device, and even if it were, it does not account for other emission sources that are not related to direct energy consumption.
As a consequence, BT is looking for a more elaborate approach which can be implemented using the data already available to incorporate all operational emissions sources into their reporting.

There are three major challenges in this matter, considering that, of course, all involved stakeholders would also like to have an as accurate accounting of their emissions as possible.
The first stems from the fact that, as discussed in Section~\ref{sec:introduction}, the guidance for applying the GHG protocol to ICT operations does not cover the case of multi-tenant data centers, such as here.
The second challenge is due to the way that the CCC service is organized, with customers using shared but off-premises private cloud infrastructure across the globe. 
We exclude for the time being from the conversation the deployment of the CCC service on public cloud services as out of scope for this study.
Because of this organization there are two apparent levels of multi-tenancy: one of BT being one of the multiple tenants in each data center used, and a second one with BT's customers being tenants of the CCC service offered by BT. 
For the purposes of the rest of this study we will be unifying these two notions under the multi-tenant term and address both of them at the same time.
In any case, the issue here is that existing approaches for calculating the carbon footprint of data centers as discussed in Section~\ref{sec:related} account for the entirety of the data center, with multi-tenancy not being sufficiently addressed.

Lastly, with the different data centers already using different amounts of green energy, and with BT purchasing RECs, as discussed above, the amount of gross emissions that have been released for the provisioning of the CCC service is different from the net difference in the amount of carbon actually released to the atmosphere due to the service.
As such we need to account for all of the above in this study.

\subsection{Study Design}

As explained in the introductory section we structure this work as an exploratory case study and follow the methodology defined by Runeson and H\"{o}rst for this purpose~\cite{runeson2009}.
As a first step we elicit requirements for a solution to the problem defined in the previous section in collaboration with stakeholders inside BT.
Through a series of meetings and after a process of consolidation and consequent updates throughout the case execution, the final list of requirements for this case is summarized in Table~\ref{tab:reqs}.
\begin{table}[t]
    \centering
    \renewcommand{\arraystretch}{1.3}
    \caption{Case requirements elicited through meetings with stakeholders at BT.}
    \label{tab:reqs}
    \scriptsize
    \begin{tabularx}{\columnwidth}{lXc}
        \toprule
         & \bfseries Requirement & \bfseries ID\\
         \cmidrule{2-3}
         \multirow{4}{*}{\bfseries Functional}
         & TCF must be calculated per tenant & FR\textsubscript{1}\\
         & The TCF calculation must include BT's Scope~1,2, and 3 emissions & FR\textsubscript{2}\\
         & A detailed breakdown of the TCF calculation must be available & FR\textsubscript{3}\\
         & A report must be automatically generated per tenant on monthly basis & FR\textsubscript{4}\\
         \midrule
         \multirow{4}{*}{\bfseries Non-Functional}
         & TCF must be distributed fairly among tenants & NFR\textsubscript{1}\\
         & The TCF calculation must be auditable & NFR\textsubscript{2}\\
         & The generated report must fit in one page, to force the contents to be concise and the information easy to consume & NFR\textsubscript{3}\\
         \bottomrule
    \end{tabularx}
\end{table}

Going through the table, it becomes clear that there are two tasks to be carried out.
First, in the absence of a suitable model for the calculation of TCF in multi-tenant data centers one has to be developed for this purpose.
The model needs to be developed in close collaboration with the case provider (BT) so as to ensure that it is actually feasible, that is, the data required for its calculation are available and accurate.
The model does not need to be complete, in the sense that we know that Scope~3 emissions are difficult to account for, but it does need to be fair to all tenants, and allow them to verify the calculations independently.
Having such a model developed allows for the second task to proceed: to develop a report generator producing monthly reports summarizing each tenant's TCF in an easy to digest manner.
This allows us to use generated reports as the means of collecting feedback from BT's customers for the purposes of evaluating our proposal.

The following sections discuss these two tasks: first covering the development of the TCF calculation model, and then embedding the development of the report generator into the wider context of evaluating the proposed model.

\section{Total Carbon Footprint Calculation Model}
\label{sec:model}

We base our model for obtaining the Total Carbon Footprint $TCF$ of a service tenant hosted in a multi-tenant data center based on the notion of the three scopes defined by the GHG protocol (Section~\ref{sec:background}).
Scope~4 as defined by Matthews et al. is not being considered, since the fine-grained information for distinguishing among the Scope~3 emissions is not available in our case.
The formula for calculating the $TCF$ is therefore:
\begin{equation}
TCF=Scope_1 + Scope_2 + Scope_3\label{frm:tcf}    
\end{equation}
\noindent where $Scope_i$ is the emissions calculated for each scope.
All components of this formula, and of all other formulas to follow in this section, are \emph{tenant-specific} except if indicated otherwise; we omit adding a tenant index in the formulas for reasons of simplicity but it can otherwise be assumed to be there.
In the following we decompose Eq.~\ref{frm:tcf} to its constituents and explain how $TCF$ can be calculated in practice.
For all energy measurements we presume that the unit used is Watt-hours, while for the emissions themselves we assume grams of CO\textsubscript{2}-equivalent emissions.

\subsection{Scopes 1 \& 3}

Before defining the formulas for $Scope_1$ and $Scope_2$ we need first to define the notion of responsibility ratio $r$ meant to represent the amount of the emissions that a particular tenant is actually responsible for.
There are two separate areas of responsibility to be considered by this work: between tenants of the same service, and between the data centers and their tenants.
Following Lenzen and Murray, we adopt the position that the emissions at data center level should be attributed with a percentage $x$ to the provider, and $1-x$ to the tenants of the data center. 
The $1-x$ remaining percentage should consider all emissions that by necessity have to be shared among tenants, such as e.g.\ the ones resulting from cooling the data center.

The formula to calculate the ratio $r$ of a tenant at a given data center, covering both areas of responsibility, is given by:
\begin{align}
    r &= \lambda \times L_{share} \label{frm:r}\\
    \lambda &= \frac{Scope_2^{tenant}}{Scope_2^{DC}} \label{frm:lamda}
\end{align}
\noindent where $L_{share}$ is the attributed share to the tenant according to the Lenzen and Murray method and $\lambda$ the ratio of the Scope~2 emissions for the tenant to the total (Scope~2) emissions for the data center ($DC$).
Since Scope~2 emissions are associated with the consumption of electricity and related activities, we effectively use this definition of $\lambda$ as a way of associating the overall responsibility ratio $r$ to that due to power consumption: the more energy ``hungry'' are the activities for a specific tenant, the bigger the share of the overall emissions attributed to them.

For the purposes of realizing this model as discussed in Section~\ref{sec:evaluation}, and in order mainly to stress the accountability of tenants in this model, we set $L_{share}$ to $100\%$, meaning that tenants are responsible for all emissions accruing from their use of the data center equipment.
This is meant to act as an upper boundary for calculating these emissions and a more in-depth method to calculate $L_{share}$ accordingly is left as future work at this point.

Given the above we define:
\begin{align}
    Scope_1 &= \sum_{DCs} \sum_{devices} F_{device} \times c_{device} \times r \label{frm:scope1}\\
    Scope_3 &= \sum_{DCs} Scope_3^{DC} \times r \label{frm:scope3}
\end{align}
\noindent where 
\begin{itemize}
    \item $F_{device}$ and $c_{device}$ the fuel consumed and amount of CO\textsubscript{2} emissions per gram of fuel consumed by the device, also known as its \emph{carbon intensity}, respectively, for all devices in each data center where the tenant resides, 
    \item $Scope_3^{DC}$ the total Scope~3 emissions of each data center that the tenant resides, to be measured directly as a set of related activities in this scope (where possible), and,
    \item $r$ the responsibility ratio of the tenant for each data center, as calculated by Eq.~\ref{frm:r}.
\end{itemize}

We therefore only need to define how $Scope_2$ emissions are to be calculated next, since we also need them for Eq.~\ref{frm:lamda} and by extension Eq.~\ref{frm:r}.

\subsection{Scope 2}

Scope~2 incorporates the emissions of the different devices operating in a data center.
The formula calculating these emissions is:
\begin{equation}
    Scope_2 = \sum_{DCs} ((E_{DC}^{server} + E_{DC}^{network} + E_{DC}^{cooling} + E_{DC}^{other} ) \times c_{DC} \times L_{share})
\end{equation}

\noindent where $E_{DC}$ is the energy consumed by the tenant's use of $server$ (including storage) and $network$ devices, the energy consumed by the $cooling$ system as a consequence of their heat production, plus any $other$ source of energy consumption such as lighting; $c_{DC}$ is the carbon intensity of the energy source (i.e.\ the electricity provider) used by the data center as a whole, and $L_{share}$ the tenant's share as discussed above. 

If the data center is owned, operated, and managed by the same organization then it might be possible to measure directly the energy consumption of each device and convert into emissions by multiplying it with the carbon intensity of the electricity mix at the point of the measurement.
Both in the general and the specific case of BT this kind of measurement is not possible, and we therefore need to provide estimates for the various $E_{DC}$ components.

There are a number of approaches in the literature for estimating the server energy component $E^{server}$\,\footnote{In the following discussion we omit writing explicitly the $DC$ index for each energy component only for brevity; it can otherwise be assumed to be used for all energy components.}, for example the methods defined by Da Silva et al.~\cite{dasilva2020}, Bohra and Chaudhary~\cite{bohra2010}, the follow up to that work by Kansal et al.~\cite{kansal2010}, etc.
These methods are basically interchangeable in our model, and a decision on which one to use depends mainly on the availability of data required for them to work.
In the case under consideration we opt for the Bohra and Chaudhary (BC, for short) method which needs only four parameters: CPU usage, cache data, DRAM data, and disk usage data, arranged in a weighted sum.
In order to find the weight for each one of these parameters we run a benchmark on each server device type as per the BC method, and build a linear model for them through linear regression.

For the network energy component $E^{network}$ there are fewer methods that can be used. 
Two particular methods are popular in the literature on this topic: the works by Aslan et al.~\cite{aslan2018} and the more sophisticated method by Mahadevan et al.~\cite{mahadevan2009}.
Since the Mahadevan et al.\ method requires energy consumption per network device port for its estimation that is simply not available in our case we use the Aslan et al.\ one instead.
This basically means that we estimate network energy consumption as \[E^{network}=6\times 10^{-8}\times(bytes_{sent} + bytes_{received})\]i.e.\ by assuming $6\times 10^{-8}$ Watt-hours consumed per byte passing through the device.
The management interface of each network device can be used to retrieve the total amount of bytes per tenant since each tenant-specific server and storage device is assigned to distinct port(s).

Energy used for cooling devices cannot easily be attributed to each tenant directly, at least not without being able to measure the temperature of each device that the tenant has some load on.
The approach we take instead is to split it according to the tenant's share of overall power consumption, similar to Scope~1 and 3 emissions.
We therefore define $E^{cooling}$ across all data centers that the tenant is using as:
\begin{equation}
    E_{DC}^{cooling} = \sum_{devices} E_{DC}^{device} \times \frac{E_{DC}^{server} + E_{DC}^{network}}{\sum_{tenants} (E_{DC}^{server} + E_{DC}^{network})} \label{frm:cooling}
\end{equation}
\noindent where $E_{DC}^{device}$ is the total power consumed by a cooling device in the data center, and with the fraction representing the ratio of directly attributable energy consumed per tenant (that is, for server and network devices) to the total energy consumed for all data center tenants for the same type of devices.
The same formula can be used also for the rest of the energy consumption sources component $E_{other}$ by simply taking into account all other devices instead of the cooling ones.

\subsection{Net Emissions}

The Total Carbon Footprint calculated by Eq.~\ref{frm:tcf} and through the respective formulas for the different scopes is effectively representing the gross emissions of each tenant.
Since, however, emission offsetting measures might be taken, as discussed for example in the case of BT purchasing RECs and of data center operators using renewable energy as part of their electricity mix, a net emissions $TCF$ needs to be defined accordingly.
For this purpose we define in addition:
\begin{equation}
    TCF_{net} = \sum_{DCs} (TCF_{DC} - E_{{DC}_{green}} \times c_{DC} - REC_{DC}\times r) \label{frm:net}
\end{equation}
\noindent where $E_{green}$ is the amount of renewable energy being used by each data center, $REC$ the amount of carbon offset through the respective certificates, $c_{DC}$ the carbon intensity for the data center as above, and $TCF_{DC}$ the footrpint of the whole data center, calculated as the sum of the $TCF$ for all its tenants.
For the evaluation of our proposal presented both the gross (Eq.~\ref{frm:tcf}) and the net TCF (Eq.~\ref{frm:net}) calculations will be used.

\section{Evaluation}
\label{sec:evaluation}

Looking at the elicited case requirements summarized in Table~\ref{tab:reqs} it becomes clear that some of them (NFRs~2 and 3) require input from actual or at least potential users of the system to be evaluated with respect to their degree of satisfaction.
Furthermore, and while FR\textsubscript{1}, FR\textsubscript{2}, and NFR\textsubscript{1} are arguably satisfied by construction through the model elaborated in the previous section, the rest of the requirements point to an implementation of the model for their satisfaction.
Consequently, in the following we discuss the development of a sustainability report generator for the BT case, and the process we followed to assess the remaining requirements.
After that we go back to the research questions defined in the introduction to discuss our findings through this case study, and in closing identify the limitations of our work.

\subsection{Report Generator}
\label{sec:generator}

Since there is no requirement for integrating our model into an existing reporting system for the purposes of this case we opted to develop one as a stand-alone proof-of-concept tool implementing the theoretical model of the previous section.
The resulting report generator was written in Kotlin, making it executable on any machine with a Java Virtual Machine installed.
The generator is built as a pipeline of three independent but interconnected steps: reading input, calculating the carbon footprint, and generating the report itself.
Each step only needs to ensure that it handles its input in an expected format and outputs a result in another expected format.
As a result, the actual implementation of each step can be easily adjusted or replaced without modifying the other components in the pipeline.
Each of these steps is discussed in more detail in the following.

\paragraph*{Reading input} The generator takes as input comma-delimited (CSV) files containing the usage information on the server and network equipment per data center, plus information about each data center in terms of e.g.\ energy consumed for cooling, and a separate file per tenant with details about the tenant (distribution of load across data centers etc.) 
The files are exported by BT's personnel using the administrative dashboards for the management of their servers and network devices, usually organized per calendar month.
Beyond some checks that all necessary information is available, not much processing takes place in this step which concludes by transforming the provided data into a \texttt{RawData} object and passing it on to the next step.

\paragraph*{Calculating the TCF} The next step in the pipeline is to take the \texttt{RawData} object and use it to calculate the carbon footprint of each tenant. 
To achieve this, this step also calculates a number of metrics required for the $TCF$ model such as the total energy consumption for all tenants in a data center, energy consumption per device type, and gross and net emissions as a whole for BT's operation of the CCC service, and per tenant.
The latter results are composed into a separate \texttt{Footprint} object per tenant.
In the current implementation each \texttt{Footprint} object represents one calendar month of a tenant's use of the CCC service.
However, the model itself is time scale-agnostic, and the TCF is calculated to match the timescale of the data provided to it.
Since the provided input is on a monthly basis then the treatment of the data logically follows suit.

\paragraph*{Generating the report} The final step in the pipeline is to generate a report for each tenant.
As the means to satisfy FRs~3 and 4, the sustainability report covering each tenant's TCF is generated into two formats based on the same \texttt{Footprint} object.
The first, and more elaborate report is made available as a JSON file, making it much easier to deal with programmatically, allowing also for the auditability of the presented results when combined with the theoretical model (NFR\textsubscript{2}).
Listing~\ref{lst:json} in~\ref{sec:appendix:listing} shows a small fragment of a generated report in JSON format for a fictitious tenant; as it can be seen in the listing, both the data used as input for each device, and the consequent calculations per scope are contained in the file.

The other, intentionally more user-friendly option towards satisfying NFR\textsubscript{3} generates the report in a single-page PDF format.
To make this part as configurable as possible we actually created a \LaTeX~template with placeholders in it to be replaced by actual values from the \texttt{Footprint} object, and relied on the LuaLaTeX compiler\footnote{\url{https://github.com/lualatex}} to generate the report after the values have been set.
Fig.~\ref{fig:report} in~\ref{sec:appendix:report} shows a generated report in PDF format for the same fictitious tenant as in the case of the JSON format.

Both formats of the report (JSON and PDF) consist of the same five sections; the only difference is that the JSON format contains in addition the detailed breakdowns for each calculation.
The first section summarizes the gross and net emissions for the tenant in a month, as well as providing an overview of the carbon emission per agent, and a comparison with the same data from the previous two months.
This section is represented by the top-left part of Fig.~\ref{fig:report}.
The second section puts the gross emissions into perspective by comparing them to the number of Amsterdam-New York flights, number of kilometers drive by an average car, and amount of smartphones charged for the same amount of emissions.
As it can be seen in the top-right part of the figure, publicly available data are used for the calculation of these equivalent emissions.

The third section of the report is a breakdown of which sources the emissions consist of, visualized in Fig.~\ref{fig:report} as a pie graph in the middle and left of the page.
This pie graph visualizes the Scope~1, 2 and 3 emissions per tenant, with the Scope~2 actually broken down into server, network, cooling, and other emissions following Section~\ref{sec:model}.
The other pie graph in the figure is summarizing the information in the fourth section of the report, showing the share of emissions offset per offset method, as well as the resulting net emissions.
At the bottom of the page there are additional details concerning the $TCF$ model and the data used for preparing this report. 

\subsection{Tenants Survey}

To collect input from potential users of the report generator we arranged interviews using separate video calls with three account holders at BT, representing some of the most heavy users of the CCC service.
In each call the overall project was briefly introduced, the report for the tenant that the account holder is representing was made available for inspection in both formats, and consequently a set of semi-closed questions were asked to determine the satisfaction level with respect to the respective requirements.
More specifically, the following questions were asked:
\begin{enumerate}
    \item How clear is the report and the method behind it?
    \item How useful is the information from both formats of the report?
    \item How appropriate do you consider the method for calculating shared emissions?
\end{enumerate}
All interviews were recorded with the consent of the participants, and the videos were analyzed to extract the responses.

With respect to the first question (clarity), the participants overall felt that both the report and the method behind it were quite clear (NFR\textsubscript{3}), but with specific parts that need further clarification.
Participant A pointed out that the offset emissions part is unclear given their assumptions on how BT is compensating for emissions.
Participant B is quoted saying ``I don't know what's good and what's bad [...] it needs grounding'' pointing to the need to provide some point of reference for the numbers in the report.
Participant C specifically asked for a Red/Amber/Green scheme for the same purpose.
The same participant shared this view as a general comment on the work: ``It needs to be quick and easy to say where we're trending''.
Putting the provided information into more context is therefore a necessary improvement for the reporting part of our proposal.

Answering the second question (usefulness), the general consensus among the participants is that while the level of detail provided by the JSON format is great to have, it might also be intimidating for people without a technical background.
When asked as a follow-up if a spreadsheet would cause a less adverse reaction they consented that this format would be much easier to consume.
Considering both formats, the overall feeling is that the provided information is very useful (NFR\textsubscript{2}).
Participant C: ``The information is of good value and it can be used, not just to improve the customer experience, but to also help the longevity of what we're doing with the customer.''
As far as the third and last question is concerned (fairness), the participants stated that the method used for dividing the emissions seems fair (NFR\textsubscript{1}).
Participant B answered: ``As long as we can describe our hypotheses and our methodology, and it's auditable and traceable, then it seems fair.''
Participant C stated that ``It's important when we have a company like [company name redacted], who use data services in Europe and Asia, but not in the Americas, we are tailoring the report to the infrastructure they use.''
Beyond these specific questions, all participants found this work both very important and timely, and were impressed by the level of detail offered.

\subsection{Findings}


We organize our findings around the research questions defined already in Section~\ref{sec:introduction}.
More specifically, to answer RQ1 (How to apply the GHG Protocol to report the TCF of tenants in multi-tenant data centers?) we discussed how the Protocol can be adapted in the case of multi-tenant data centers hosting multi-tenant software services.
The focus in our case study was mostly on Scope~2 emissions since there are no significant grounds for Scope~1 emissions, as discussed further later in this section, and Scope~3 emissions are not as easy to obtain.
However, the presented model accounts for both Scope~1 and~3 emissions, nonetheless.
Furthermore, any carbon footprint offsetting methods need to be also taken into account and subtracted from the TCF calculation for estimating net, instead of gross emissions.

RQ2 (How can the TCF of a data center be accurately obtained?) was answered through the $TCF$ model developed in Section~\ref{sec:model}.
Where accurate monitoring data through e.g.\ power measuring devices are available these data can be used directly in the model; where this is not possible the model offers the means to estimate them instead.

Finally, with respect to RQ3 (How can the TCF be fairly distributed over all tenants of a multi-tenant data center?) we observe that assigning responsibility for emissions in a shared computing environment is a largely unsolved and not much researched problem.
We based our approach to the one by Lenzen and Murray and we therefore suffer from the same shortcoming in arbitrarily setting the $L_{share}$ for each party, opting to assign it to $100\%$ on the tenant side for the purposes of this work.
Beyond this point requiring future work, server and network devices in our $TCF$ model can have their energy consumption calculated per tenant according to each tenant's use of the device.
For shared devices for which it is not possible to trace each tenant's individual usage, the energy consumption, and through that, carbon emissions can be divided according to the same percentage of the total server and network device consumption the tenant is responsible for.
Calculating the TCF for each tenant then requires aggregating their $TCF$ as calculated for each data center which hosts the service that the tenant is consuming.

A point that needs to be made here is about the choice of using Scope~2 as the basis of proportionally allocating tenant responsibility since it affects the definition of $\lambda$ in Eq.~\ref{frm:lamda} and by extension the responsibility ratio $r$ (Eq.~\ref{frm:r}).
Data center operators are, for example, contractually obliged to run their backup electricity generators once per month for maintenance purposes, contributing therefore to Scope~1 emissions.
This only lasts however for a short period of time (usually around 30' per month). 
As such, Scope~1 emissions are at least an order of magnitude smaller in comparison to Scope~2 emissions, at least in our described case, and are accounted anyway through the definition of $Scope_1$. 
Gupta et al.~\cite{gupta2021} also use data from Google and Facebook to show that Scope~3 emissions have overtaken Scope~2 emissions in the recent years in data center operations.
However, it is unclear how their findings apply to cloud service providers that are not at the scale of Google and Facebook in terms of number of data centers, i.e.\ the vast majority of providers outside of the hyperscalers, including the case described here.
We therefore cannot reliably add them to the calculation of the responsibility ratio, at least not without further investigation.

\subsection{Limitations}
\label{sec:limitations}

While we tried to make the model as accurate and complete as possible, there are still several limitations mainly in its implementation through the report generator.
The first limitation emerges from the treatment of carbon intensity as a constant value in the timespan of each generated report, i.e.\ a month.
In reality, carbon intensity is better defined as a function over time, especially if we consider the use of renewable energy sources such as solar panels at data center level and how their output varies during the span of a day.
Despite the model needing an extension in this direction, it also needs to be considered that since most of the load to the data centers is probably during office hours when also the output of these sources is at its highest then using an average carbon intensity over the whole month is an acceptable solution.

The other limitation stemming from the model is our use of a linear model for the calculation of the weights necessary for estimating the energy consumption per server device.
For the data we had available for our implementation, our regression was tested with an adjusted $R^2=0.3626$, meaning that it has limited reliability.
A more sophisticated regression method, and access to more data over longer periods of time, can help us develop more accurate weights for this calculation.
In addition, our model in its current form leans on the assignment of each of BT's service tenants to distinct server devices to treat a device shared by two tenants as essentially two separate devices, each occupied by one tenant.
To remedy this situation and allow for sharing also of the same server device the energy consumption model for servers needs to be updated to estimate first the consumption of the entire device and then use the ratio between the tenants' estimated one to divide the device's emissions.
This, together with the development of a more appropriate model for setting the $L_{share}$ responsibility value as discussed in Section~\ref{sec:model} is left for future work.

The most important shortcoming of this work though is, without a doubt, its lack of model validation.
Performing this validation empirically would require collecting tenant-specific energy consumption data from the data center operator(s) involved, aggregate them, convert them into carbon footprint through the carbon intensity of the data center energy sources, and compare them against the respective calculations by the $TCF$ model.
The step that makes this process inapplicable in practice, however, is exactly collecting the tenant-specific data. 
Even if instrumentation were available for each server and network device that the tenant is using, and at the same time these devices were dedicated to the single tenant, there is still the by definition shared devices to consider, such as the HVAC equipment cooling down the data center as a whole. 
At this point in time we therefore argue for the validity of the model by construction, and look for opportunities to validate it empirically through e.g.\ a large-scale experiment.

Finally, and perhaps more obviously, our approach was designed with specific requirements in the context of the case study with BT, and having their cloud infrastructure in mind.
Using our approach requires a level of transparency from the data center operator which makes it unsuitable in the case of less privileged relationships between tenants and data center operators, and definitely not suitable for public cloud deployments.
A different model, such as the one proposed by ThoughtWorks, as discussed in the following section, might be a better fit in the last case, but further research is required, especially to deal with cases of hybrid cloud deployments.

\section{Related Work}
\label{sec:related}

The related work on the topic can be separated into two categories: industry-driven efforts and academic research.
With respect to the former, as discussed in the introductory section, the three hyperscalers already offer carbon footprint monitoring dashboards to their customers since at least last year.
All three are offering breakdowns per region and service, in addition to showing the total footprint of the attributed to the customer emissions over time.
On one hand, offering these dashboards has motivated a number of smaller providers and even software vendors such as VMWare to add sustainability dashboards\footnote{See \url{https://blogs.vmware.com/management/2021/10/sustainability-dashboards-in-vrealize-operations-8-6.html} for an example, even though a point needs to be made that VMWare started incorporating sustainability concerns before the hyperscalers made their dashboards available to the public.} to their offerings.
At the same time, however, the methodology behind the development of these dashboards is unclear or too vague when it comes to providing details on how emissions are measured or estimated and subsequently attributed to individual tenants, and they are provider- or vendor-specific to boot, making them unsuitable for adoption by this work.

Beyond these dashboards, a number of open source tools and projects have been created both for specific providers and solutions, and for platform-agnostic purposes.
An example of the former is Cloud Foundry Footprint\footnote{\url{https://github.com/P-Ehlert/cloudfoundry_footprint} (Accessed May 8, 2023)}, essentially a Python script that estimates the carbon footprint of Cloud Foundry applications based on a simple estimation model.
With respect to the latter, approaches like the CarbonFootprintCalculator tool\footnote{\url{https://github.com/MarosMacko/CarbonFootprintCalculator} (as above)} or the spreadsheets available on the web site of the GHG Protocol itself\footnote{\url{https://ghgprotocol.org/Tools_Built_on_GHG_Protocol} (as above)} allow for emissions calculation without considering the details of technology involved.
However, they require manual input of data that beyond making them cumbersome, is in some cases simply not possible due to lack of visibility in the operations of a data center.
As a compromise, ThoughtWorks offers their own open source carbon footprint calculator\footnote{\url{https://www.cloudcarbonfootprint.org/} (as above)} for primarily cloud service providers which takes as input monthly bills and outputs estimates based on a model of emissions for each provider.
While this makes this tool very easy to adopt for cloud consumers, and can estimate emissions for private cloud deployments, it does presume that one of the hyperscalers is being used for input purposes and is therefore not generic enough.

In terms of academic research, relevant approaches have focused on measuring, estimating, or modeling either energy consumption (and through that carbon footprint) or directly emissions at data center level.
Falling in the first category, Bohra and Chaudhary~\cite{bohra2010}, for example, proposed the method used by our model which takes into account CPU, memory, hard disk and network usage and combines them using weights differing per workload.
Smith et al.~\cite{smith2012} developed CloudMonitor, an approach for automatically obtaining these weights.
Kansal et al.~\cite{kansal2010} created a similar model and automatic calculation method which they incorporated into Joulemeter, a tool which integrates with Microsoft's \mbox{Hyper-V} hypervisor to monitor VM data directly.
Lin et al.~\cite{lin2018} built upon these approaches to propose DEM, an automated power consumption estimation tool.
The review of the existing methodologies in measuring power consumption on data center level conducted by Vasques et al.~\cite{vasques2018} concluded, however, that more research is needed in this area as there are multiple issues with them.

On a different tack, it is worth mentioning the work by Jagroep et al.~\cite{jagroep2016} which proposes the use of a Resource Utilization Score (RUS) instead of the more commonly used PUE (Power Usage Effectiveness) as a way of gaining fine grain visibility into the effect of individual components of a data center to the emissions.
Lastly, estimating the power consumption due to network usage at and through data centers is still an open research area, with the survey by Aslan et al.~\cite{aslan2018} showing wide deviations between the estimated amounts for different models.
As a result, the work by Mahadevan et al.~\cite{mahadevan2009} and the formula it defines for network equipment has become a very popular method for this purpose.
Tackling directly data center emissions, Belkhir and Elmegigi~\cite{belkhir2018} present a deep dive in both the embodied and operational  emissions of individual devices in a data center.
The survey of Gill and Buyya~\cite{gill2018} offers a wide overview of the current state of the art in both power consumption and emissions measurement and management for cloud data centers.

Of particular note, the data center-focused chapter in the GHG Protocol ICT guidance~\cite{ghgguidance} discusses a methodology which follows closely the GHG Protocol to propose two methods of measuring emissions: bottom-up and top-down. 
Bottom-up means that individual emissions are calculated for small parts of the data center, for example individual devices, and then summed up to get the total emissions.
Top-down means that the total emissions are calculated first, and then divided over individual physical and virtual machines,  services and users. 
The emissions are divided according to the percentage  of the data center capacity that is assigned to each machine, service, and user.
Our proposal is therefore a bottom-up approach according to this categorization.
Ultimately, all these approaches with the partial exception of the last one, while useful on their own, suffer from the same deficit that this work addresses: their primary audience is data center operators and not SaaS providers, and as a result do not account neither for the multi-tenancy of cloud data centers nor for that of the software services hosted in these centers.

\section{Conclusions}
\label{sec:conclusions}

The previous sections presented our proposal on allocating the carbon footprint in terms of operational emissions to tenants of SaaS services hosted in multi-tenant (that is, shared) data centers.
This model was developed in the context of an exploratory case study done in collaboration with BT Global Services acting as a SaaS provider for virtualized call center services to companies around the world.
Primary goals for this model is fairness in the way it distributes emissions between tenants, and transparency in the way that it does so.
Given the lack of a suitable model which allows for multi-tenancy to be taken into consideration we proceeded to develop one ourselves building on the GHG Protocol and existing work.
We then discussed how a report generator can be implemented to realize this model using two different formats for the presented information, and used generated reports to ask potential adopters of our proposal for feedback.
The input we received from them provides evidence that our model is perceived overall very positively, with improvements required in the future in how the information is communicated to users rather than the model itself.

In terms of further future work, our ongoing and planned efforts aim at addressing the limitations discussed in Section~\ref{sec:limitations}, and in particular with respect to incorporating a variable in smaller time increments carbon intensity values, developing more sophisticated models for the sharing of responsibility between the data center(s) and tenants and between tenants sharing the same server devices, and collecting data for a more reliable regression model when to comes to server energy consumption estimation.
Furthermore, automating the delivery of the necessary input data to the report generator pipeline discussed in Section~\ref{sec:generator} which at this point is still being done manually, will make the overall process more accurate and less labor intensive.

\section*{Acknowledgements}

The authors would like to thank Christos Bantis for his invaluable contribution to this work.

\clearpage
\appendix
\section{Example generated report}
\label{sec:appendix:listing}
\begin{lstlisting}[frame=single,language=JSON,showstringspaces=false,basicstyle=\scriptsize\ttfamily, caption=Fragment of a generated TCF report for a fictitious BT tenant showing the breakdown per scope., captionpos=t, floatplacement=h,label=lst:json]
"scopes": {
        "scope1": {
          "type": "pipeline.datatypes.Scope1",
          "isAggregate": false,
          "energy": 0.0,
          "emissions": 0.0
        },
        "scope2": {
          "type": "pipeline.datatypes.Scope2",
          "isAggregate": false,
          "energy": 4500000,
          "emissions": 1800000,
          "devices": {
            "servers": {
              "SERVER_1234": {
                "type": "pipeline.datatypes.ServerDevice",
                "isAggregate": false,
                "deviceModel": "ABC_987",
                "energy": 100000.0,
                "emissions": 40000.0,
                "utilization": 0.10,
                "cacheMoved": 2e7,
                "dramAccessed": 5e9,
                "diskMoved": 2e10
              },
            },
            "network": {
              "NETWORK_DEVICE_1234": {
                "type": "pipeline.datatypes.NetworkDevice",
                "isAggregate": false,
                "deviceType": "router",
                "energy": 100000.0,
                "emissions": 40000.0,
                "bytesSent": 1e12,
                "bytesReceived": 1e12
              },
            }
\end{lstlisting}

\clearpage
\section{Example generated single-page report}
\label{sec:appendix:report}
\begin{figure*}[h]
\centering
\includegraphics[width=.8\linewidth]{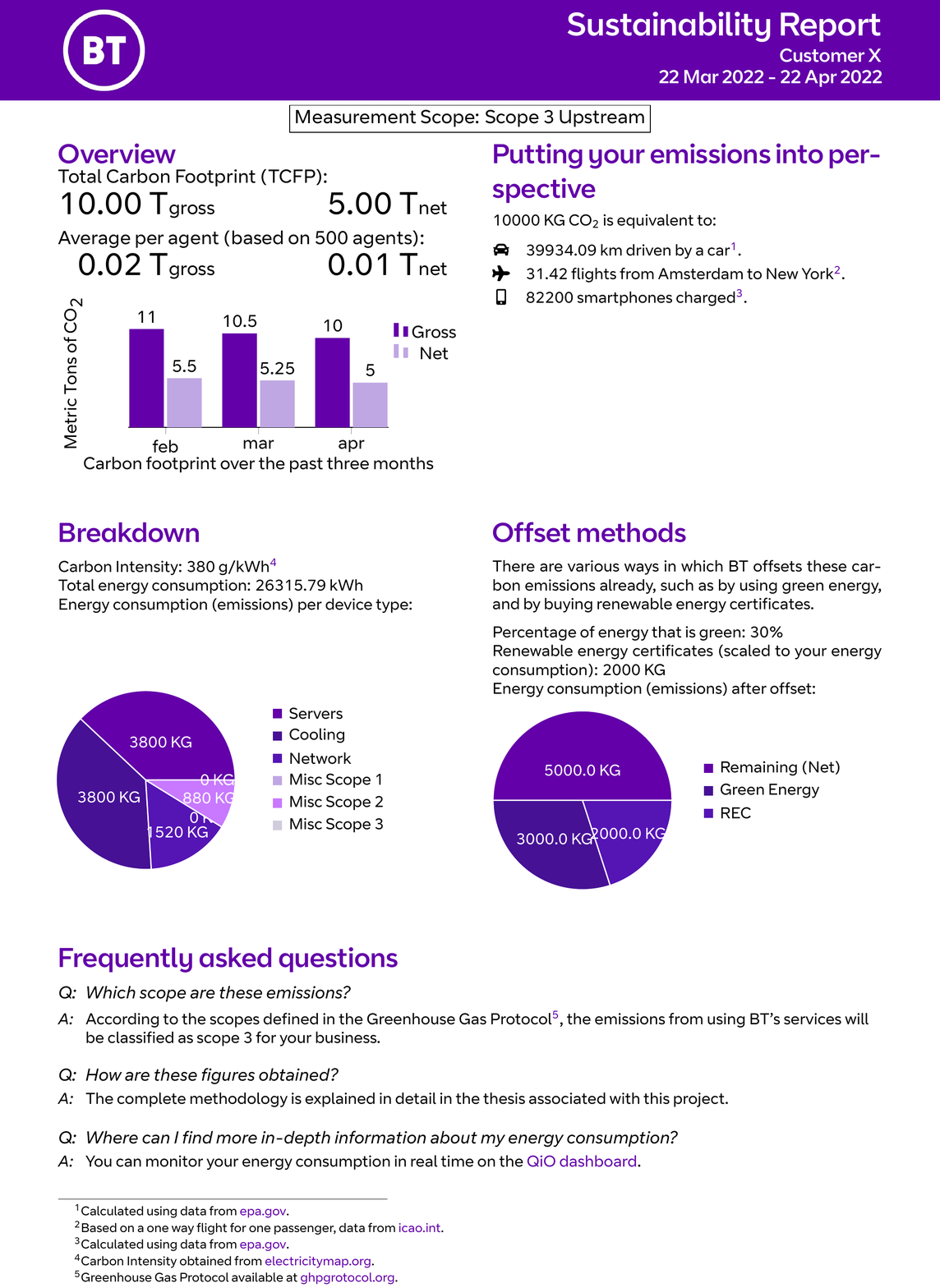}
\caption{Sample one-page TCF report generated for the same fictitious customer.}
\label{fig:report}
\end{figure*}
\clearpage

\bibliographystyle{elsarticle-num} 
\bibliography{TCF_allocation_model_arxiv}





\end{document}